\documentclass[border=20pt]{article}
\usepackage[a4paper, total={6in, 8in},margin=1.5cm]{geometry}
\usepackage{amsmath,amssymb,amsthm, amsfonts,latexsym}
\usepackage[T1]{fontenc}
\usepackage[english]{babel}
\usepackage[utf8]{inputenc}
\usepackage{mathtools}
\usepackage[numbers,sort&compress]{natbib}
\usepackage[breaklinks,hidelinks]{hyperref}
\usepackage{cleveref}
\usepackage{authblk}
\usepackage{comment}
\usepackage{lineno}
\usepackage{breqn}

\usepackage{algorithm}
\usepackage{algpseudocode}

\usepackage{float,afterpage}
\usepackage{blkarray}

\title{An Algebraic Approach to Evolutionary Accumulation Models}

\author[1,*]{Jessica Renz}
\author[2]{Frederik Witt}
\author[1,3]{Iain G.\ Johnston}

\date{}

\affil[1]{Department of Mathematics, University of Bergen, Bergen, Norway}

\affil[2]{Department of Mathematics, University of Stuttgart, Stuttgart, Germany}

\affil[3]{Computational Biology Unit, University of Bergen,  Bergen, Norway}
\affil[*]{corresponding author, Jessica.Renz@uib.no}

\begin{document}

\maketitle

\begin{abstract}
 We present an algebraic approach to evolutionary accumulation modelling (EvAM). EvAM is concerned with learning and predicting the order in which evolutionary features accumulate over time. Our approach is complementary to the more common  optimisation-based inference methods used in this field. Namely, we first use the natural underlying polynomial structure of the evolutionary process to define a semi-algebraic set of candidate parameters consistent with a given data set before maximising the likelihood function. We consider explicit examples and show that this approach is compatible with the solutions given by various statistical evolutionary accumulation models. Furthermore, we discuss the additional information of our algebraic model relative to these models.
\end{abstract}


%
%
%
\section{Introduction}
\label{sec_introduction}
Many evolutionary processes of interest in natural sciences or medicine involve the accumulation and loss of binary features. Traditionally most used in cancer research \cite{schwartz_evolution_2017, beerenwinkel_cancer_2015,takeshima_accumulation_2019,moen_hyperhmm_2023, diaz-uriarte_every_2019}, evolutionary accumulation models have also been applied to a wide spectrum of other processes like antimicrobial resistance (AMR) \cite{nichol_steering_2015,tan_hidden_2011,greenbury_hypertraps_2020,moen_hyperhmm_2023,renz_evolutionary_2025,renz_flexible_2025} or the evolution of mitochondria \cite{johnston_evolutionary_2016,maier_massively_2013}. Furthermore, they have been used to understand and predict symptoms in progressing diseases \cite{dalgic_mapping_2021,johnston_precision_2019}, animal tool use \cite{johnston_data-driven_2020}, or learning behaviour in online courses \cite{peach_understanding_2021}. The models seek to determine the order in which certain features of interest occur, and how the presence of some features affects the probability of others appearing.\\

To address questions regarding basic knowledge about such processes or support for decision-making, several statistical models were developed during the last decades. 
Many of them are either based on a maximum-likelihood estimation (MLE) approach \cite{pagel_detecting_1994,lewis_likelihood_2001,johnston_hypercubic_2025,beerenwinkel_evolution_2006,beerenwinkel_conjunctive_2007,beerenwinkel_markov_2009,nicol_oncogenetic_2021,beaulieu_identifying_2013,beaulieu_detecting_2016,boyko_generalized_2021,schill_modelling_2020,renz_flexible_2025}, or an Expectation-Maximisation (EM) inference \cite{gerstung_quantifying_2009,moen_hyperhmm_2023,angaroni_pmce_2022,johnston_evolutionary_2016,greenbury_hypertraps_2020,aga_hypertraps-ct_2024,jahn_tree_2016}. TreeMHN \cite{luo_joint_2023} combines both and uses a regularised MLE and a hybrid Monte Carlo EM algorithm for the parameter estimation. A more detailed overview of the existing models for inferring parameters in models of accumulation or loss of binary features can be found in \cite{diaz-uriartePictureGuideCancer2025a,schill_reconstructing_2024}. All these models can fundamentally be represented by Markov models over a hypercubic state space, and involve a collection of transition parameters, as described in  \cite{johnston_hypercubic_2025}.\\

All these models are based on optimisation methods, and most of them deliver only one set of estimated parameters that describe the process. Some Bayesian methods like HyperTraPS-CT \cite{aga_hypertraps-ct_2024} use sampling to estimate a probability distribution on the transition parameters. The results depend on numerical simulation and, therefore, are not guaranteed to yield consistent results for finite computational time. Furthermore, although many of the methods provide options like bootstrapping, there is no guarantee that the solution found is a global maximum and not just a local one. Due to the high computational cost, most of these models are limited in the number of features they can handle. \\

In this article, we introduce HyperALG, an approach that describes the algebraic structure of the set of possible transition parameters given a cross-sectional dataset and neglecting sampling noise. Doing so, we shift the perspective from finding a high-likelihood solution via optimisation to a characterisation of the algebraic structure of the whole set of possible parameters. The algebraic nature that underlies these evolutionary accumulation models leads to a set of non-linear polynomial equations in the transition parameters. These polynomials are the generators of an ideal $I$, whose corresponding algebraic variety $V(I)$ contains all possible solutions as a subset $V\subseteq V(I)$. The coefficients of the polynomials can be derived from a given dataset. It is then possible to optimise the likelihood function on this algebraic variety $V$ as well. In passing, we note that this approach is scalable to any number of features. \\

 On the other hand, concrete computations using Gr\"obner bases and elimination theory \cite{cox_ideals_2018} are usually inefficient due to their enormous computational complexity. Though we rather take a theoretical model-building perspective, we note that numerical algebraic geometry could provide an efficient tool to put HyperALG into practice, see for instance~\cite{andrewjsommeseNumericalSolutionSystems2005}. In particular, homotopy continuation methods could be used to numerically approximate the solutions \cite{chenHomotopyContinuationMethod2015,verscheldeAlgorithm795PHCpack1999,BHSW06} as also discussed in Section \ref{sec_discussion}. This includes both finding individual and isolated solutions, as well as characterising positive-dimensional varieties with the help of witness sets. Further, homotopy continuation methods have proven to be efficient and robust and are now used in many different areas of application \cite{duff2023polynomial}. In addition, Hauenstein \& Sottile also introduced alphaCertified in 2012, a mathematically founded software package for certifying numerical solutions of polynomial systems \cite{hauensteinAlgorithm921AlphaCertified2012}.\\

The article is organised as follows: In the first section of this article, we will derive the defining relations of the ideal $I$ for the case of three features ($L=3$) based on the underlying evolutionary process. We will then discuss the shape of the corresponding variety $V(I)$ before showing how this setup can be generalised to situations with an arbitrary number of features. In the second part, we demonstrate our approach by concrete examples and compare the results with those of various statistical models. Finally, we discuss limitations but also further directions in which this approach could be a useful and interesting first step.

%
%
\section{Generating Equations}
\label{sec_model_equations}

%
\subsection{Evolutionary Accumulation Models}
\label{subsec_EAM}
The underlying structure of evolutionary accumulation models we want to describe is a hypercube graph of dimension $L$ whose nodes are represented by binary strings of length $L$. Each position in a binary string represents one of $L$ considered features, and a one in a string indicates the presence of the feature corresponding to this position, whereas a zero indicates the absence.\\

\begin{figure}
	\centering
	\includegraphics[scale = 0.65]{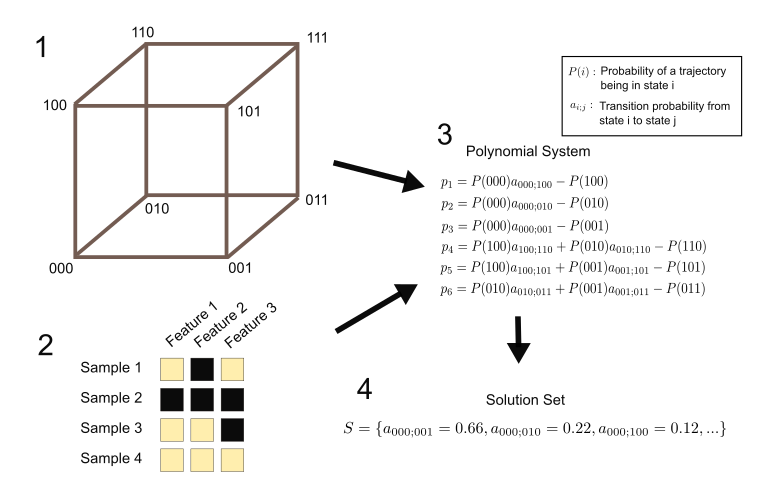}
	\caption{\textbf{HyperALG workflow. (1)} The structure of the underlying evolutionary process can be represented by trajectories on a hypercube. Here we consider three features. An evolutionary trajectory always starts in 000 and moves along the edges towards 111. The aim is to determine the transition probabilities for the different edges. \textbf{(2)} A dataset that contains information about the presence or absence of the considered features in different samples. A yellow square indicates the presence of the feature, a black square the absence. \textbf{(3)} The probability of a trajectory to pass a certain node of the (hyper)cube can be translated into a system of polynomial equations in the transition probabilities. The coefficients of the polynomials are determined by the dataset. \textbf{(4)} The zero locus of the polynomial system consists of the possible values for the transition parameters. }
	\label{overview}
\end{figure}

The process of interest is the evolutionary accumulation of these $L$ features, where we assume that every evolutionary history started at some point from the state $0^L$ in which none of the features under consideration were present. Then all features are gained one by one during this evolutionary process, which subsequently leads to the hypercube structure (see Fig. \ref{overview} (1) for $L=3$). This can be modelled by trajectories always starting at the initial node $0^L$, and proceeding stepwise along the edges of the hypercube graph into the direction of the terminal node $1^L$. Here, the Markov condition applies, i.e., the probability of which feature will be obtained next depends only on the current state, not on the trajectory's history. \\

Furthermore, we assume irreversibility, i.e., as soon as a feature is acquired, it cannot get lost again in future evolutionary steps. In particular, this imposes an ordering on the directionality of the graph. This might simplify the real situation of some evolutionary processes. Nevertheless, it is a common assumption that almost all evolutionary accumulation models follow. For the moment, only the hypercubic Mk-model \cite{johnston_hypercubic_2025} has the option to take reversibility into account, but only for a very limited number of features. The reason for this is that the irreversibility assumption makes it possible to model all trajectories in a finite number of $L+1$ steps. If, on the other hand, irreversibility is permitted, it is possible for trajectories to move arbitrarily often between intermediate states, which can lead to an infinite number of steps on the hypercube. This assumption of finiteness is also crucial for the possibility of describing evolutionary accumulation processes in an algebraic manner, as in this article. Furthermore, Aga et al. \cite{agaNaturalHistoryAMR2025a} were able to demonstrate in synthetic case studies that the assumption of irreversibility also for reversible features does not dramatically challenge the results.\\

Moreover, in most biological setups, the existence of one or more features can affect the probability of gaining or losing others. This is the reason why we consider the transition probability for each edge $i\rightarrow j$ separately from the remaining ones.  If there were just one base rate for obtaining each feature, the transition parameter would be the same for several edges, depending only on the feature that is obtained along the corresponding edge. The fact that combinations of already obtained features can increase or decrease the probability of obtaining others makes the transition parameters not only dependent on the new feature, but also on the node $i$ where the edge $i\rightarrow j$ starts.\\

%
\subsection{Underlying algebraic structure}
A directed hypercube graph itself is already an interesting object from the algebraic perspective because it defines a poset structure on the nodes, i.e., all possible evolutionary states in our case. This poset structure (the Boolean lattice) offers connections to algebraic statistics via lattice ideals \cite[ch. 1]{drtonLecturesAlgebraicStatistics2009}, and to geometric combinatorics through associated polytopes \cite{stanleyTwoPosetPolytopes1986}.\\

We exploit this structure by interpreting the evolutionary process itself as an algebraic object. Considering the transition parameters along the edges as variables, the probabilities corresponding to a specific trajectory are described by a monomial (the product of the transition parameters along the trajectory). The transition probability between arbitrary nodes (not necessarily neighbours) can then be described by polynomials -- the sum of the probabilities of all contribution paths. As we will see, we can use this to build a polynomial correspondence between a given dataset and the evolutionary process on the hypercube.\\

This intrinsic algebraic description of the process allows us to formulate conditions on the parameters to be compatible with a specific dataset. These conditions are given by polynomial equations and generate an ideal $I$, whose variety $V(I)$ contains all parameter sets that are compatible with the data.

%
\subsection{Datasets}
\label{subsec_datasets}
A dataset consists of a list of binary strings of length $L$ which represent cross-sectional data. Every string represents one sample. Again, the presence of a certain feature is indicated by a `1' at the corresponding position in the string, while a `0' indicates the absence of this feature (see Fig. \ref{overview}(2)).\\

All samples are assumed to be independent of each other and are formed by the evolutionary process under consideration. This also means that the samples share the same transition rates between different subsets of features. The purpose of evolutionary accumulation models is to learn the transition parameters. Knowing them makes it possible to do further investigations that help to understand these dependencies.\\

However, we are not primarily interested in finding a maximum-likelihood estimate, but rather in describing and characterising the space of all tuples of transition parameters compatible (as described below) with a given dataset by looking at the variety $V(I)$ for a generating ideal $I$. For this, we consider the proportions of states observed  in a finite sample of size $n$ (the dataset) as if they were the proportions observed in an infinite sample. That means that no departure from those proportions is possible, and all sampling noise is neglected. We then use these proportions to characterise a probability distribution of trajectories passing through a certain node. This probability distribution is considered separately for each evolutionary time-step, so that the sum of probabilities for all nodes with the same number of `1's in their strings sums to one. To derive this probability distribution, we also take into account that all samples from our cross-sectional dataset must have run through nodes with fewer features before, and continue to pass through further nodes with more features in the future before reaching the final node $1^L$. These probability distributions are then incorporated into our generating polynomials in the form of coefficients.\\

%
\subsection{Generating ideal for the case of three features}
\label{subsec_case_three}
In the following, we systematically define the ideal $I$, whose corresponding variety $V(I)$ contains the possible transition parameters as a subset of $\mathbb{R}^{10}$, see~\eqref{general_V}. We are developing $I$ in several steps, considering the three different types of structures that are reflected in the generators of $I$: 1. polynomials describing the process dynamics (Equation \eqref{orig_equations}), 2. polynomials taking into account the probability normalisation (Equation  \eqref{P(j)_sum_to_1}), and 3. polynomials that guarantee trajectory consistency (Equation \eqref{trajectory_consistency}).\\

First, we want to do this for the case of three considered features, i.e., the string length $L=3$. The main reason for restricting ourselves to this low number of features is that a smaller system leads to fewer generators of $I$ as well as a smaller dimension of the polynomial ring in which $I$ is defined. Consequently, it is easier to handle and interpret $V(I)$, and to understand how $I$ is derived. However, all the steps described below can be equally well applied to hypercube transition graphs of higher dimensions, cf.\ Subsection \ref{subsec_generalization}.\\

Let $P(i)$ be the proportion of trajectories in the considered evolutionary process that pass via node $i$, where $i \in \{000,001,010,011,100,101,110,111\}$, and $a_{i;j}$ the transition parameter from node $i$ to node $j$. Because the step we are interested in is the characterisation of the compatible transition parameters $a_{i;j}$, we consider them as variables. Then
\begin{equation}\label{def_p(j)}
	P(j) = \sum_{i} P(i)\cdot a_{i;j}\in \mathbb{C}[a_{i;j}]_{i,j}.
\end{equation}
Taking into account the directed hypercube structure, in this sum occur only those terms for $i$ that are directly connected to $j$ by an edge and whose string contains a smaller number of `1's. All other summands disappear, for we must have $a_{i;j} = 0$ by design. This reduces the number of variables $a_{i;j}$ of interest to the number of edges, which is $2^{L-1}L$ in general, and 12 in our case, where $L=3$.\\

Furthermore, the sum of the $P(i)$ for all $i$ with the same number of `1's in it must be 1, because $P(i)$ describes a proportion of all trajectories for every evolutionary step. In addition, the condition that 
\begin{equation}\label{a_ij_sum_to_one}
	\sum_{j} a_{i;j} =1
\end{equation}
must hold true for every $i$.\\

Equation \eqref{def_p(j)} gives us an equation for a proportion that we can assign to every node, each having the shape of a polynomial in the variables $a_{i;j}$. As there is only one equation per node, we get a first ideal $I_{1}$ which describes the process dynamics and is generated by $2^{L}$ polynomials. For instance, we get the following generators for $I_{1}\subseteq\mathbb{C}[a_{000;100},a_{000;010},a_{000;001},a_{001;101},a_{001;011},a_{010;110},a_{010;011},a_{100;110},a_{100;101}]$ in the case of $L=3$:
\begin{equation}\label{orig_equations}
	\begin{split}
		f_{1}=&1- P(000),\\
		f_{2}=&P(000)a_{000;100} - P(100),\\
		f_{3}=&P(000)a_{000;010} - P(010),\\
		f_{4}=&P(000)a_{000;001}- P(001),\\
		f_{5}=&P(100)a_{100;110} + P(010)a_{010;110} - P(110),\\
		f_{6}=&P(100)a_{100;101} + P(001)a_{001;101}- P(101),\\
		f_{7}=&P(010)a_{010;011} + P(001)a_{001;011} - P(011),\\
		f_{8}=&1 - P(111).
	\end{split}
\end{equation}
The polynomials taking care of the probability normalisation give rise to a second ideal 
\[ 
	I_{2} \subseteq \mathbb{C}[a_{000;100},a_{000;010},a_{000;001},a_{001;101},a_{001;011},a_{010;110},a_{010;011},a_{100;110},a_{100;101}]
\]
given by
\begin{equation}\label{P(j)_sum_to_1}
	\begin{split}
	I_{2} = ( f_{9}&,f_{10},f_{11},f_{12},f_{13},f_{14}) = ( P(100)+P(010) + P(001) - 1,P(110) + P(101) + P(011) - 1,\\ & a_{000;100} + a_{000;010} + a_{000;001} -1, a_{100;110} + a_{100;101} -1,a_{010;110} + a_{010;011} -1,a_{001;101} + a_{001;011} - 1).
	\end{split}
\end{equation}
The variables $a_{110;111}, a_{101;111}$ and $ a_{011;111}$ all have to be 1, independent of the considered dataset, so we don't need to take them further into account.\\

Taking the information from both $I_{1}$ and $I_{2}$ into account, we can reduce the number of variables even further without loosing information and consider the ideal $I_{1} \cup\, I_{2}$ in the polynomial ring $ \mathbb{C}[a_{000;100},a_{000;010},a_{001;101},a_{010;110},a_{100;110}].$ This ideal is initially generated by 14 polynomials, the eight generators from $I_{1}$ together with the six generators from $I_{2}$. However, an analysis shows that there exist non-trivial syzygies between those generators: 
\[
\left(1,1,1,1,0,0,0,0,1,0, -P(000),0,0,0\right) \text{ and } \left(0,0,0,0,1,1,1,0,-1,1,0,-P(100),-P(010),-P(001)\right).
\]
 This reduces the generating system to 12 polynomials and points to a rich intrinsic algebraic structure. Note that the reduction to exactly those five variables is a particular choice of a generating set. One could equally well choose other generators in another set of free variables for the same ideal, which would describe the same structure, although considered in another subring.\\

Next, we derive polynomials for the $P(i)$, $i\notin\{000,111\}$ that connect them to the dataset. Namely, every trajectory that is in a certain position at the moment the sample is taken necessarily had to pass through previous nodes of the hypercube before. We therefore introduce additional variables $b_{i;j}$ for every edge $i\rightarrow j$, which describe the ratio of the trajectories in $j$ that passed via $i$ before. Again, it holds that
\begin{equation*}
	b_{010;110} = 1-b_{100;110}, \quad b_{001;011} = 1 - b_{010;011}, \quad b_{001;101} = 1-b_{100;101},
\end{equation*}
and 
\begin{equation*}
	b_{110;111} = 1-b_{101;111} - b_{011;111},
\end{equation*}
as well as 
\begin{equation*}
	b_{000;001} = b_{000;010} = b_{000;100} = 1.
\end{equation*}
 Furthermore, the internal relationships between the original $a_{i;j}$ and the additional $b_{i;j}$ variables yield to a further ideal 
\[
	I_{3}\subseteq\mathbb{C}[a_{001;101},a_{001;011},a_{010;110},a_{010;011},a_{100;110},a_{100;101},b_{001;101},b_{001;011},b_{010;110},b_{011;111},b_{101;111}]  
\]
guaranteeing trajectory consistency. It is generated by
\begin{equation}\label{trajectory_consistency}
	\begin{split}
		& b_{001;101}\left(P(001)a_{001;101} + P(100)a_{100;101}\right) - P(001)a_{001;101},\\
		& b_{001;011}\left(P(001)a_{001;011}+P(010)a_{010;011}\right) - P(001)a_{001;011},\\
		&b_{010;110}\left(P(010)a_{010;110}+P(100)a_{100;110}\right) - P(010)a_{010;110},\\
		&b_{011;111}\left(P(101) + P(110) + P(011)\right) - P(011),\\
		&b_{101;111}\left(P(101)+P(110)+P(011)\right) - P(101).
	\end{split}
\end{equation}
\\

Finally, we determine the proportions of each state from the dataset by defining
\begin{equation}\label{N_i_easy}
	N_{i} := \frac{D_{i}}{n}
\end{equation}
for each node in the hypercube, where $D_{i}$ is the absolute number of state $i$ in the dataset and $n$ the total number of data points. Ultimately, the $P(i)$, that is, the proportion of trajectories passing via node $i$, acquire the subsequent polynomial expressions in $\mathbb{C}[a_{i;j},b_{i;j}]_{i,j}$ with coefficients determined by the parameters $N_l$:
\begin{equation}
	\label{Pi_equations}
	\begin{split}
		P(100) = &N_{000}a_{000;100} + N_{100} + N_{110}(1-b_{010;110}) + N_{101}(1-b_{001;101}) + N_{111}(1-b_{001;101})b_{101;111} \\ &+ N_{111}(1-b_{010;110})(1-b_{011;111}-b_{101;111}),\\
		P(010) = &N_{000}a_{000;010} + N_{010} + N_{110}b_{010;110} + N_{011}(1-b_{001;011}) + N_{111}b_{010;110}(1-b_{011;111} - b_{101;111})\\& + N_{111}(1-b_{001;011})b_{011;111},\\
		P(001) = &N_{000}(1-a_{000;010}-a_{000;100}) + N_{001} + N_{101}b_{001;101} + N_{011}b_{001;011} \\ &+ N_{111}b_{001;101}b_{101;111} + N_{111}b_{001;011}b_{011;111},\\
		P(110) =& N_{000}a_{000;100}a_{100;110} + N_{000}a_{000;010}a_{010;110} + N_{100}a_{100;110} + N_{010}a_{010;110} + N_{110} \\&+ N_{111}(1-b_{011;111} - b_{101;111}),\\
		P(101) =& N_{000}(1-a_{000;010} - a_{000;100})a_{001;101} + N_{000}a_{000;100}(1-a_{100;110}) + N_{001}a_{001;101} \\&+ N_{100}(1-a_{100;110}) + N_{101} + N_{111}b_{101;111},\\
		P(011) =& N_{000}(1-a_{000;010} - a_{000;100})(1-a_{001;101}) + N_{000}a_{000;010}(1-a_{010;110}) \\ &+ N_{001}(1-a_{001;101}) + N_{010}(1-a_{010;110})+ N_{011} + N_{111}b_{011;111}.
	\end{split}
\end{equation}
As an example, the expression of $P(100)$ is obtained from the proportion of trajectories $N_{000}a_{000;100}$ that are sampled in $000$ and will move to $100$, the proportion of trajectories $N_{100}$ reflecting the trajectories directly sampled in $100$, and the proportion of trajectories $N_{110}b_{100;110} = N_{110}(1-b_{010;110})$ sampled in $110$ originating from $100$ etc.\\

Combining all information and substituting the equations \eqref{Pi_equations} for the corresponding $P(001),P(010),\ldots$ we finally obtain our generating ideal $I= I_{1}\cup\, I_{2} \cup \, I_{3} \in \mathbb{C}[a_{i;j},b_{i;j}]_{i,j}$ which is generated by the following list of polynomials: 
\begin{equation*}
	\begin{split}
		& a_{000;100} - P(100),\\
		& a_{000;010} - P(010),\\
		& P(100)a_{100;110} + P(010)a_{010;110} - P(110),\\
		& P(100) (1- a_{100;110}) + P(001)a_{001;101} - P(101),\\
		& b_{001;101}\left(P(001)a_{001;101} + P(100)(1-a_{100;110})\right) - P(001)a_{001;101},\\
		& b_{001;011}\left(P(001)(1-a_{001;101})+P(010)(1-a_{010;110})\right) - P(001)(1-a_{001;101}),\\
		&b_{010;110}\left(P(010)a_{010;110}+P(100)a_{100;110}\right) - P(010)a_{010;110},\\
		& (P(101)+P(110)+P(011)) b_{011;111} - P(011),\\
		& (P(101)+P(110)+P(011)) b_{101;111} - P(101).
	\end{split}
\end{equation*}
The nine generators are non-linear polynomials $p_{i}(a_{k;j},b_{k;j}), i \in \{1,\ldots,9\}$ in the variables $a_{k;j},\ b_{k;j}$, with coefficients $N_\ell$ given by the dataset. \\
At this point, we would like to emphasise that the introduction of the variables $b_{i;j}$ is something that is new compared to the existing evolutionary accumulation models, which only use the transition parameters $a_{i;j}$  that appear in the likelihood function in order to model the trajectories over the hypercube. It may seem counterintuitive to introduce new additional variables and thereby drastically increase the dimension of the polynomial ring. An illustrative example of why this is nevertheless necessary can be found in Appendix \ref{example_nec_b}.

%
\subsection{Parameter set as subset of the real locus of $V(I)$}
\label{subsec_solution_set}
The parameter set we are interested in is a real subset of the algebraic variety $V(I)$. In general, considering $V(I)$ as a complex variety, we can also encounter non-real solutions or real solutions outside the interval $[0,1]$. These do not, of course, represent possible solutions to our problem, as the parameters represent proportions. Among all real points in $V(I)$, those with $a_{i;j},b_{i;j}\in [0,1]$ give us the transition probabilities we wanted to learn from the dataset. Schematically, our solution set is therefore the semi-algebraic set
\begin{equation}
\label{general_V}
	V =  V(I) \cap \mathbb{R}^{10} \cap [0,1]^{10},
\end{equation}
where the exponent 10 corresponds to the number of variables $(a_{000;100},a_{000;010},a_{001;101},a_{010;110},a_{100;110},b_{001;101},\allowbreak b_{001;011},b_{010;110},b_{011;111},b_{101;111})$.\\

From a statistical point of view, the elements in $V\subseteq V(I)$ may have different likelihood values. As described in Subsection \ref{subsec_datasets}, we consider the proportions of states given by the dataset as if they were the proportions observed in an infinite sample, which makes deviations from these proportions impossible. Furthermore, the transition parameters are not considered as probabilities but as relative proportions in a closed system so that the number of considered individuals remains constant. Therefore, $V$ does not contain only the parameters that most likely lead to the observed data, but all tuples of transition parameters that are consistent with the data, considering a closed system. However, it is possible to find the maximum likelihood estimate by optimising the likelihood function over $V$.\\

%
\subsection{Generalisation to an arbitrary number of features}
\label{subsec_generalization}
The approach deployed in \ref{subsec_case_three} for the case of three features can be generalised to an arbitrary number of considered features. For a general string of length $L$, we get one polynomial generator for $I_{1}$ for every one of the inner $2^L -2$ nodes in the hypercube, in accordance with the equations in \eqref{def_p(j)}. Removing the redundant generators, this number can be reduced to $2^L - 2- (L-1) = 2^L - L -1$.\\

The proportions for the zero-string $P(0^L)$ and the one-string $P(1^L)$ stay 1 in all cases. As in \eqref{P(j)_sum_to_1}, $\sum_{i} P(i) =1$ for all $i$ that contain the same number of `1' in the binary string. Also, Equation \eqref{a_ij_sum_to_one} and with that the generating system of $I_{2}$ can be transferred one-to-one to the general case.\\

The number of free transition variables $a_{i;j}$ is given by
\begin{equation*}
	\sum_{k = 0}^{L-2} \frac{L!}{k!(L-k)!}(L-1-k).
\end{equation*}
Additionally, we get
\begin{equation*}
	\sum_{k = 2}^{L}\frac{L!}{k!(L-k)!}(k-1)
\end{equation*}
variables $b_{i;j}$ for taking into account `future' states, and each of these variables gives rise to a generating polynomial of $I_{3}$ 
\begin{equation*}
	b_{i;j}\left(\sum_{k,|k| = |i|}P(k)a_{k;j}\right) - P(i)a_{i;j}
\end{equation*}
reflecting the inner dependencies between the $b_{i;j}$ and the $a_{i;j}$ variables.\\

The only generalisation that is slightly more involved is the polynomial expressions that connect the $P(i)$ to the dataset. To see how the pattern works, we first look at the polynomial descriptions in the case of $L=4$.  We then have to distinguish between three different evolutionary time-steps. For the nodes in step one that have only one `1' in their binary string ($|i|=1$), we can calculate $P(i)$ from the polynomial
\begin{equation*}
	\begin{split}
		P(i)_{|i|=1} =& N_{i} + \sum_{|k| = |i| -1} N_{k}a_{k;i} +  \sum_{|k|=|i|+1}N_{k}b_{i;k} +  \sum_{|l|=|i|+2}\sum_{|k| = |i|+1} N_{l}b_{i;k}b_{k;l} \\&+ \sum_{|m| = |i|+3} \sum_{|l|=|i| +2} \sum_{|k|=|i|+1} N_{m}b_{i;k}b_{k;l}b_{l;m}\\
		=& N_i + N_{0000}a_{0000;i} +\sum_{|k|=2}N_{k}b_{i;k} + \sum_{|l|=3}\sum_{|k|=2}N_{l}b_{i;k}b_{k;l} \\&+ \sum_{|l|=3}\sum_{|k|=2}N_{1111}b_{i;k}b_{k;l}b_{l;1111}.
	\end{split}
\end{equation*}
Namely, we have to consider only one previous step, but two steps afterwards. That leads to the double and triple sum. Here, we assume that all $a_{i;j}$ and $b_{i;j}$ where $i$ and $j$ are not directly connected by an edge $i\rightarrow j$ in the hypercube are set to zero. Consequently, the corresponding monomials disappear.\\

When considering the states one step further, that is, $|i|=2$, we have to consider up to two previous steps but only one future step. This leads to the double sum over the $a_{i;j}$ variables in the polynomial
\begin{equation*}
	\begin{split}
		P(i)_{|i|=2} =& N_{i} + \sum_{|l|=|i|-2} \sum_{|k| = |i|-1} N_{l}a_{l;k}a_{k;i} + \sum_{|k| = |i|-1}N_{k}a_{k;i} + \sum_{|k|=|i|+1}N_{k}b_{i;k} \\&+ \sum_{|l|=|i|+2} \sum_{|k|=|i|+1} N_{l}b_{i;k}b_{k;l}\\
		=& N_i + \sum_{|k|=1} N_{0000}a_{0000;k}a_{k;i} + \sum_{|k|=1}N_{k}a_{k;i} + \sum_{|k|=3}N_{k}b_{i;k}\\ & + \sum_{|k|=3}N_{1111}b_{i;k}b_{k;1111}.
	\end{split}
\end{equation*}
In the last step, where $|i|=3$, we are in a situation where we have to go back up to three steps. This, in turn, leads to a triple sum in the variables $a_{i;j}$. On the other hand, we have only one sum that involves states from the future:
\begin{equation*}
	\begin{split}
		P(i)_{|i|=3} =& N_{i} + \sum_{|m|=|i|-3}\sum_{|l|=|i|-2}\sum_{|k|=|i|-1}N_{m}a_{m;l}a_{l;k}a_{k;i} + \sum_{|l| = |i|-2}\sum_{|k|=|i|-1}N_{l}a_{l;k}a_{k;i} \\&+ \sum_{|k|=|i|-1}N_{k}a_{k;i} + \sum_{|k| = |i|+1}N_{k}b_{i;k}\\
		=&N_{i} + \sum_{|l|=1}\sum_{|k|=2}N_{0000}a_{0000;l}a_{l;k}a_{k;i} + \sum_{|l|=1}\sum_{|k|=2}N_{l}a_{l;k}a_{k;i} + \sum_{|k|=2}N_{k}a_{k;i} \\&+ N_{1111}b_{i;1111}.
	\end{split}
\end{equation*}
Summarising, this scheme shows how to generalise the polynomials for the $P(i)$s from $L=3$ to $L=4$ by considering more steps in both directions, which eventually leads to higher multiple sums. The generalisations for $L>4$ follow the same scheme and can be obtained from the pseudo-code in Algorithm \ref{pseudo-code}.\\

\begin{algorithm}
	\caption{}
	\label{pseudo-code}
	\textbf{Generate the polynomials that connect the $P(i)$ to the dataset}
	
	\textbf{Input:} $L$= number of features, i.e., dimension of the hypercube, $N_{i}$ = proportions in the dataset of each node $i$
	
	\textbf{Output:} Polynomials $P(i)$ describing the proportion of trajectories passing node $i$
	
	\begin{algorithmic}[1]
		\State Initialise $incoming$ as a vector of length $2^L$ (entry $n$ contains indices of all edges ending at node $n$)
		\State Initialise $outgoing$ as a vector of length $2^L$ (entry $n$ contains indices of all edges starting from node $n$)
		\State Initialise $CFN$ as an empty vector of length $2^L$ (collecting the `forward coefficients' for each node)
		\State Initialise $CBN$ as an empty vector of length $2^L$ (collecting the `backward coefficients' for each node)
		\State $CFN[0] = N_{0}$; $CBN[2^L - 1] = N_{2^L-1}$
		\State $P(0) = P(2^L) = 1$
		\For {$i = 1:L-1$}
			\For {$n = 1:2^L-2$}
				\State $num$ = number of ones in node $n$
				\If {$num == i$} (Forward recursion)
					\State $CFN[n] = N_{n}$
					\For {$j$ in $incoming[n]$}
						\State $s$ = node that $j$ is starting from 
						\State $CFN[n] += CFN[s]a_{s;n}$
					\EndFor
				\EndIf	
				\If {$num == L-i$} (Backward recursion)
					\State $CBN[n] = N_{n}$
					\For {$j$ in $outgoing[n]$}
						\State $g$ = node that $j$ ends at
						\State $CBN[n] += CBN[g]b_{n;g}$
					\EndFor
				\EndIf			
			\EndFor
		\EndFor
		\For {$n=1:2^L-2$} (Combine forward and backward contributions to get $P(n)$)
			\State $P(n) = CFN[n] + CBN[n] - N_{n}$ ($N_{n}$ is contained in forward and backward contribution, i.e., we have to remove one)
		\EndFor
		\algstore{myalg1}
	\end{algorithmic}
	
	\vspace{0.3cm}
	\textbf{Generating polynomials of ideal $I_1$}
	
	\textbf{Input:} $L$ = number of features, i.e., dimension of the hypercube, $incoming, \ outcoing$ and the $P(i)$ from above
	
	\textbf{Output:} Vector $G_{1}$ containing the generating polynomials of $I_1$
	
	\begin{algorithmic}[1]
		\algrestore{myalg1}
		\State Initialise $G_1$ as an empty vector of length $2^L-2$ for collecting the generators of $I_{1}$
		\For {$n = 1:2^L -2$}
			\State $G_1[n] = -P(n)$
			\For {$j$ in $incoming[n]$}
				\State $s$ = node that $j$ is starting from 
				\State $G_1[n] += P(s)a_{s;n}$
			\EndFor
		\EndFor
		\algstore{myalg2}
	\end{algorithmic}
	
	\vspace{0.3cm}
	\textbf{Generating polynomials of ideal $I_2$}
	
	\textbf{Input:} $L$ = number of features, i.e., dimension of the hypercube, $outgoing$ and the $P(i)$ from above
	
	\textbf{Output:} Vector $G_{2}$ containing the generating polynomials of $I_2$
	
	\begin{algorithmic}[1]
		\algrestore{myalg2}
		\State Initialise $G_2$ as an vector of length $L-1$ containing $-1$ in each coordinate
		\For {$n = 1:2^L-2$}
			\State $step$ = number of ones in node $n$
			\State $G_{2}[step] += P(n)$
		\EndFor
		\For {$n = 0:2^L -1$}
			\If {number of ones in node $n \neq L-1$}
				\State $gen = -1$
				\For {$j$ in $outgoing[n]$}
					\State $g$ = node that $j$ ends at
					\State $gen += a_{n;g}$
				\EndFor
				\State push!($G_{2},gen$)
			\EndIf
		\EndFor
	\algstore{myalg}
	\end{algorithmic}
\end{algorithm}

\begin{algorithm}
	\textbf{Generating polynomials of ideal $I_3$}
	
	\textbf{Input:} $L$ = number of features, i.e., dimension of the hypercube, $incoming$ and the $P(i)$ from above
	
	\textbf{Output:} Vector $G_{3}$ containing the generating polynomials of $I_3$
	
	\begin{algorithmic}[1]
		\algrestore{myalg}
		\State Initialise $G_3$ as an empty vector for collecting the generators of $I_3$
		\For {$n=1:2^L-1$}
			\If {number of ones in $n \neq 1$}
				\State $sum = 0$
				\For {$j$ in $incoming[n]$}
					\State $s$ = node that $j$ is starting from 
					\State $sum += P(s)a_{s;n}$
				\EndFor
				\For {$j$ in $incoming[n]$}
					\State $s$ = node that $j$ is starting from
					\State $ poly = -P(s)a_{s;n} + sum\cdot b_{s;n}$
					\State push!($G_{3},poly$)
				\EndFor
			\EndIf
		\EndFor
	\end{algorithmic}
\end{algorithm}

%
\subsection{Computational realisation}
\label{sec_software}
The code for the generation of the polynomials for an arbitrary $L$ and a given dataset is implemented in Julia \cite{Julia-2017} and can be found on Github: \url{https://github.com/JessicaRenz/HyperALG}. It requires the Julia packages OSCAR \cite{OSCAR,OSCAR-book} and DelimitedFiles. For finding approximations of solutions for the polynomial systems in Subsection \ref{subsec_ovarian}, we used the Julia package Optim \cite{Optim.jl}. Within this package, the Broyden-Fletcher-Goldfarb-Shanno algorithm (BFGS) was used with the default arguments. The initial values for all variables were set to 0.5 for obtaining the values for $b_{i;j}$ for the statistical models.\\

Both the number of $a_{i;j}$ and the number of $b_{i;j}$ variables grows in $\mathcal{O}(L\cdot 2^L)$. Regarding the number of generators, the number of polynomials generating $I_1\cup I_2$ grows in $\mathcal{O}(2^L)$, and the number of polynomials generating $I_3$ grows also in $\mathcal{O}(L\cdot 2^L)$ since every variable $b_{i;j}$ comes with a generator for $I_3$. The degree of the polynomials grows in $\mathcal{O}(L)$. In addition, the computation of a Gr\"obner basis can be doubly exponential in the number of variables in the worst case scenario, which becomes quickly unwieldy.

%
%
\section{Results}
\label{sec_results}
In this section, we illustrate our approach with a toy dataset for $L=3$ and a real medical dataset for $L=4$ and compare our results with those of the statistical models HyperLAU \cite{renz_flexible_2025}, HyperTraPS \cite{aga_hypertraps-ct_2024} and HyperHMM \cite{moen_hyperhmm_2023} and discuss the pros and cons of our algebraic approach in comparison to the statistical models. 

%
\subsection{Example and comparison with statistical models for $L=3$}
\label{subsec_L3Exam}
We consider the toy dataset
\[
D=[D_{000},D_{001},D_{010},D_{011},D_{100},D_{101},D_{110},D_{111}] = [0, 2, 0, 0 , 1, 5, 0,0],
\]
cf.\ Equation~\eqref{N_i_easy}. A Gr\"obner basis for our ideal 
\[
	I\subseteq\mathbb{C}[a_{000;100},a_{000;010},a_{001;101},a_{010;110},a_{100;110},b_{001;101},b_{001;011},b_{010;110},b_{011;111},b_{101;111}]
\]
with respect to grevlex order (for an explanation  of monomial orders see, for example, \cite[p. 58]{cox_ideals_2018}) is given by the following list of polynomials:  
\begin{equation*}
	\begin{split}
		&3b_{001;101} + 16b_{011;111} + 8b_{101;111} -10,\\
		&a_{100;110} + 8b_{011;111} + 8b_{101;111} - 8,\\
		& 2a_{001;101} - a_{100;110} - 8b_{101;111} + 6,\\
		& a_{000;010},\\
		&8a_{000;100} + 5b_{001;101} -6,\\
		&8b_{101;111}^2 - b_{011;111} - 15b_{101;111} + 7,\\
		& 16b_{011;111}b_{101;111} + 8b_{101;111}^2 - 13b_{011;111},\\
		&b_{011;111}^2 + b_{011;111}b_{101;111} - b_{011;111},\\
		& b_{010;110}b_{011;111} + b_{010;110}b_{101;111} - b_{010;110},\\
		&b_{001;011}b_{011;111} - b_{011;111},\\
		&b_{001;011}b_{010;110}b_{101;111} - b_{001;011}b_{010;110} - b_{010;110}b_{101;111} + b_{010;110}.
	\end{split}
\end{equation*}
Then $\dim(I) = 3$, and $I$ leads to the corresponding variety
\begin{equation*}
	\begin{split}
	V(I) = & \left\{\left(\frac{1}{3},0,1,a_{010;110},0,\frac{2}{3},b_{001;011},b_{010;110},0,1\right)\right\}\cup \left\{\left(\frac{1}{8},0,1,a_{010;110},1,1,b_{001;011},0,0,\frac{7}{8}\right)\right\}\\
	&\cup \left\{\left(\frac{3}{4},0,0,a_{010;110},0,0,1,b_{010;110},\frac{1}{4},\frac{3}{4}\right)\right\},
	\end{split}
\end{equation*}
with $a_{010;110},b_{001;011},b_{010;110} \in \mathbb{C}$.
From an applied point of view, we are mostly interested in the values for $a_{i;j}\in[0,1]$ as long as there exist corresponding values $b_{i;j}\in [0,1]$ so that the point $(a_{i;j},b_{i;j})\in V\subseteq V(I)$. We obtain $V$ by restricting all free variables in $V(I)$ to the real interval $[0,1]$.\\

Next, we want to compare this result with results that are obtained by several of the currently available statistical evolutionary accumulation models. Since the statistical models only recognise the $a_{i;j}$-variables, we define the following projection
\begin{equation*}
	\pi: V \rightarrow V_{\text{proj}}: \mathbb{R}^{10} \rightarrow \mathbb{R}^{5}
\end{equation*}
projecting on the first five coordinates:
\begin{equation*}
	\begin{split}	\pi(a_{000;100},&a_{000;010},a_{001;101},a_{010,110},a_{100;110},b_{001;101},b_{001;011},b_{010;110},b_{011;111},b_{101;111})\\ &= (a_{000;100},a_{000;010},a_{001;101},a_{010;110},a_{100;111}).
	\end{split}
\end{equation*}
Here, this projection gives rise to the following image:
\begin{equation*}
	V_{\text{proj}} = \left\{\left( \frac{1}{3},0,1,a_{010;110},0\right)\cup \left( \frac{1}{8},0,1,a_{010;110},1\right)\cup \left(\frac{3}{4},0,0,a_{010;110},0\right)\middle|\, a_{010;110}\in [0,1] \right\}.
\end{equation*} 
\\

Feeding the input data $D$ to HyperLAU \cite{renz_flexible_2025}, we obtain the solution set
\begin{align*}
	S_{\text{HyperLAU}} =& \{(a_{000;100},a_{000;010},a_{001;101},a_{010;110},a_{100;110})\} =  \{(0.33 , 0,1,0.6, 0) \},
\end{align*}
which is actually a subset of $V_{\text{proj}}$.\\

This also holds for the solutions from HyperTraPS-CT \cite{aga_hypertraps-ct_2024} and HyperHMM \cite{moen_hyperhmm_2023}, which are 
\begin{align*}
	S_{\text{HyperTraPS}} = &\{(0.335, 0, 1, 0.999, 0)\}
\end{align*}
and 
\begin{align*}
	S_{\text{HyperHMM}} = &\{(0.334, 0, 1, 0.5,0)\},
\end{align*}
respectively.\\

\begin{figure}
	\centering
	\includegraphics[scale=0.65]{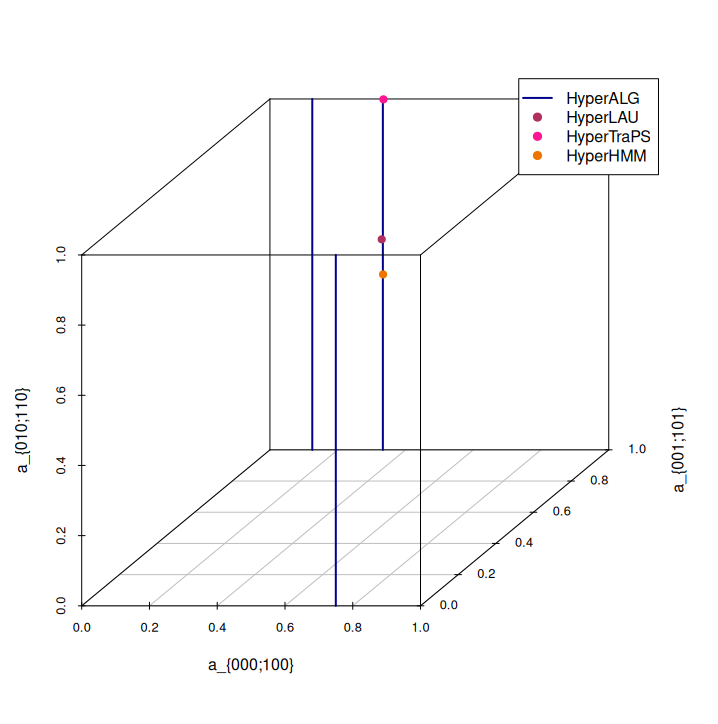}
	\caption{\textbf{Solutions in the unit cube.} Reported solutions of the different approaches, projected to three dimensions representing the parameters $a_{000;100},a_{001;101}$ and $a_{010;110}$. All three statistical models (HyperLAU, HyperTraPS and HyperHMM) find a solution (illustrated as points) on the same component reported by HyperALG (illustrated as lines).}
	\label{3D}
\end{figure}

Summarising, these models yield a point in the same component of $V_{\text{proj}}$, but choose one particular value for the non-identifiable variable $a_{010;110}$ corresponding to one particular point in the set $V_{\text{proj}}$ and are in alignment with HyperALG, which reports all of them as possible solutions (see also Figure \ref{3D}).\\

Furthermore, for every point $p \in V_{\text{proj}}$, i.e., for every solution given by a statistical model that is compatible with the evolutionary system and the given data, the projection $\pi$ defines for us the fiber $\pi^{-1}(p) \subseteq V\subseteq V(I)$. This fiber indicates all the possible $b_{i;j}$-values that fit the given transition-probabilities. For the solution for HyperLAU, for example, the fiber is
\begin{equation*}
	\pi^{-1}(S_{\text{HyperLAU}}) = \left\{\left( 0.33,0,1,0.6,0,\frac{2}{3},b_{001;011},b_{010;110},0,1\right)\middle|\, b_{001;011},b_{010;110} \in [0,1]\right\}.
\end{equation*}
As we can see here, even when choosing a concrete point $p \in \pi(V) = V_{\text{proj}}$, the fiber $\pi^{-1}(p)$ can still be positive dimensional, i.e., we have variables that are non-identifiable. This illustrates an interesting biological interpretation, that many different evolutionary `histories' (here represented by the $b_{i;j}$ variables) can be connected with the same forward evolutionary dynamics (given by the $a_{i;j}$).\\

Looking now on the likelihood function $\mathcal{L}$ for sampling this dataset, we get
\begin{equation*}
	\begin{split}
	\mathcal{L} \propto&   \prod_{|i|=1} a_{000;i}^{N_{i}} \cdot \prod_{|i|=2}\left( \sum_{|j|=1}a_{000;j}a_{j;i}\right)^{N_{i}} \\
	=& \;a_{000;001}^{2} \cdot a_{000;100}^{1} \cdot (a_{000;100}a_{100;101} + a_{000;001}a_{001;101})^{5}\\
	=& (1-a_{000;100}-a_{000;010})^{2} \cdot a_{000;100}^{1} \cdot (a_{000;100}(1-a_{100;110})\\
	&+(1-a_{000;100}-a_{000;010})a_{001;101})^{5}.
	\end{split}
\end{equation*} 
Restricting this function to $V$ yields
\begin{equation*}
	\begin{split}
	\mathcal{L} &= \left(1-\frac{1}{3} - 0\right)^{2} \cdot \frac{1}{3} \cdot \left(\frac{1}{3}\left(1-0\right)+\left(1-\frac{1}{3}-0\right)\cdot 1\right)^5 \\
	&= \left(\frac{2}{3}\right)^{2} \cdot \frac{1}{3} \cdot \left(\frac{1}{3}+\frac{2}{3}\right)^{5} = \frac{4}{27} \approx 0.15,
	\end{split}
\end{equation*}
or
\begin{equation*}
		\mathcal{L} = \left(1-\frac{1}{8}-0\right)^2\cdot \frac{1}{8} \cdot \left( \frac{1}{8}\cdot (1-1)+\left(1-\frac{1}{8} - 0\right)\cdot 1 \right)^5 \approx 0.05,
\end{equation*}
or 
\begin{equation*}
	\mathcal{L} =  \left(1-\frac{3}{4} - 0\right)^2\cdot \frac{3}{4}\cdot \left(\frac{3}{4}\cdot (1-0) + \left(1-\frac{3}{4} - 0\right)\cdot 0 \right) ^ 5 \approx 0.01,
\end{equation*}
depending on which component of $V$ is considered. The statistical models find a point that lies on the first component, giving the best likelihood, as expected.\\

In general, since the likelihood function only depends on the $a_{i;j}$-variables, all points lying on the same fiber have the same likelihood. In this example, we can even consider a family of fibers $\pi^{-1}_{a_{010;110}}$, which is parameterised by the non-identifiable variable $a_{010;110}$, and contains all $\pi^{-1}(S_{\text{HyperLAU}}),\ \pi^{-1}(S_{\text{HyperTraPS}})$, and $\pi^{-1}(S_{\text{HyperHMM}})$. Since $a_{010;110}$ doesn't appear in the likelihood function, all points in this whole family of fibers give rise to the same likelihood value.\\

Note that the likelihood function itself is a polynomial in any dimension. However, the local or global optimum can be located at the boundary of the unit cube, in which case it can't be detected by the partial derivatives of the likelihood function.

%
\subsection{Ovarian cancer data with $L=4$}
\label{subsec_ovarian}
Next, we apply HyperALG to a subset of a cancer dataset \cite{Knutsen_2005}, which has already been used to test several evolutionary accumulation models \cite{greenbury_hypertraps_2020,szabo_estimating_2002,moen_hyperhmm_2023,loohuis_2014}. Originally, the presence or absence of seven different chromosomal aberrations in 87 samples of ovarian cancer patients was reported. Here, we focus on a subset of these mutations and consider the following four features: $8q+$, $3q+$, $5q-$, and $4q-$. These labels indicate the index of the chromosome where the mutation occurs, as well as the chromosomal arm ($p$ or $q$). A $+$ indicates addition while $-$ denotes a loss.\\

For $L=4$ features we get 28  generating polynomials of the ideal $I$ in 34 variables ($17 \times a_{i;j}$ and $17\times b_{i;j}$). Unless $I$ is of quite simple shape, we cannot expect to get an explicit parametrisation of $V(I)$ as in the case of a linear system. Also, the computation of the whole variety, as we have done in Subsection \ref{subsec_L3Exam}, is not feasible on a reasonable time scale. To approximate at least some of the points on $V$, we use the Julia package Optim \cite{Optim.jl} to minimise the function $\sum_{i=1}^{28} p_{i}^{2}$, where $p_{i}$, $i = 1,...,28$ are the polynomial generators obtained by following the procedure described in Section \ref{sec_model_equations} (using different initial parameter values for the optimisation). Some selected parameter tuples received can be found in Table \ref{variable_values} in Appendix \ref{appendix_results}. \\

To compare our approach with some statistical models, also in this case, we run HyperLAU \cite{renz_flexible_2025}, HyperHMM \cite{moen_hyperhmm_2023}, and HyperTraPS \cite{aga_hypertraps-ct_2024} to obtain their solutions for the $a_{i;j}$ values. Since these models do not provide values for the $b_{i;j}$, we substitute the $a_{i;j}$-values into the polynomials and obtain a new system that now depends on the variables $b_{i;j}$ only. We minimise again the sum of all squared polynomials by using the Optim package and make sure that we use values between 0 and 1. The estimated parameters can be found in Table \ref{variable_values} in Appendix \ref{appendix_results}, too.\\

After substitution of the obtained values into the system of the 28 polynomials, we would expect values equal to or close to zero (due to numerical deviations). Although the values depend on the selected approach, we can see (cf.\ Table \ref{derivations}, Appendix \ref{appendix_results}) that this is indeed the case for all reported solutions. In particular, we consider the solutions of the statistical models to be close to our variety $V \subseteq V(I)$.\\

Finally, we compare the likelihood of the different solutions we obtained. The general likelihood function for $L=4$ is proportional to
\begin{equation*}
	\mathcal{L} \propto  \prod_{|i|=1} a_{0000;i}^{N_{i}} \cdot \prod_{|i|=2}\left( \sum_{|j|=1}a_{0000;j}a_{j;i}\right)^{N_{i}} \cdot  \prod_{|i|=3}\left(\sum_{|j|=2}\sum_{|k|=1}a_{0000;k}a_{k;j}a_{j;i}\right)^{N_{i}}.
\end{equation*}
For our concrete ovarian cancer dataset from \cite{Knutsen_2005} we have $N_{1} = 2, \ N_{2} = 3,\ N_{3} = 5,\ N_{4} = 2,\ N_{5} = 1,\ N_{6} = 2,\ N_{7} = 1,\ N_{8} = 7,\ N_{9} = 3,\ N_{10} = 1,\ N_{11} = 8,\ N_{12} = 12,\ N_{13} = 4 \text{ and } N_{14} = 6$.\\

For convenience, we consider the log-likelihood $\ell = \log(\mathcal{L})$ instead of the likelihood values themselves. They are given by
\begin{equation*}
	\begin{split}
	\ell_{\text{HyperALG}_1}&= -74.19,\ \ell_{\text{HyperALG}_2} = -74.21,\  \ell_{\text{HyperALG}_3} = -78.96,\\ 
	 \ell_{\text{HyperHMM}} &= -74.06,\ 
	  \ell_{\text{HyperLAU}} = -74.06, \ \ell_{\text{HyperTraPS}} = -76.33.
	\end{split}
\end{equation*}
We see, that the likelihoods are broadly in the same range, but also that there are points with different likelihood values on the variety given by HyperALG, as expected by design.\\

Summarising, this demonstrates that the solutions provided by the three statistical models are compatible with the variety defined by HyperALG. Furthermore, we find other points on $V$ of equal quality from a likelihood perspective. Note, however, that we didn't compute the complete variety $V$. In particular, it is conceivable that there exist additional solutions that yield an equally good likelihood.

%
%
\section{Discussion}
\label{sec_discussion}

We present a new approach describing compatible tuples of transition parameters in an evolutionary accumulation model for a given dataset as points of a semi-algebraic set $V\subseteq V(I)$. The generating polynomials are derived directly from the dynamics of the underlying biological process. This framework can be applied to any number of features and provides the basis for a systematic analysis of the algebraic structure of the parameter space. \\

In dimension three, we have illustrated how the 10-dimensional parameter space projects naturally onto a 5-dimensional subspace, determining the likelihood. This projection induces a fiber structure, where points in the same fiber have identical likelihoods. \\

Based on the construction of this set, we would expect that the maximum likelihood estimate is included in it. Despite the lack of a general formal proof, we were able to verify this for all examples considered. Of course, due to numerical approximations and rounding, it cannot be expected that all deviations will be exactly zero, and minor divergences such as those we see here naturally arise. \\

The advantage of the HyperALG approach lies in its algebraic nature. In principle, the characterisation of the compatible parameters via an ideal and a semi-algebraic set enables one to use symbolic tools from algebra, such as Gr\"obner basis computation and elimination theory, to determine the solution set and to analyse algebro-geometric characteristics like the dimension. However, the computational complexity of these methods prevents a systematic use for HyperALG polynomials, even for relatively small numbers of features. \\

One possible line of attack is to use methods from numerical algebraic geometry, which provide tools to compute points on the variety using homotopy continuation (for example Bertini \cite{BHSW06}, PHCpack \cite{verscheldeAlgorithm795PHCpack1999,verscheldePolynomialHomotopyContinuation2011} or HomotopyContinuation.jl \cite{HomotopyContinuation.jl}). However, even with these advanced methods, practical applicability remains limited: for the systems considered in this paper, the resulting homotopy computations still exhibit prohibitively long runtimes.\\

Apart from the efficient calculation of the algebraic variety, there are a whole series of further interesting questions on the theoretical side: What is the generic dimension of the variety as a function of the model dimension, i.e., the number of considered features? When do particular data patterns cause jumps or degeneracies in dimension or degree? Under what conditions of the data does HyperALG have (real) solutions? Do distinct irreducible components of the variety correspond to biologically different evolutionary scenarios? Does the algebraic structure give systematic insights into identifiability?\\

For now, we hope that this approach can serve as an interesting starting point for developing new ideas and advancing the field of applied computational algebra and mathematical biology.

%
%
\section*{Acknowledgements}
This work was supported by the Trond Mohn Foundation [project HyperEvol under grant agreement No. TMS2021TMT09 to IGJ], through the Centre for Antimicrobial Resistance in Western Norway (CAMRIA)[TMS2020TMT11].

\bibliographystyle{plainnat}
\bibliography{used_only}
\clearpage
\appendix

%
%
\section{Necessity of considering earlier and later states}
\label{example_nec_b}
The following example illustrates what problems can occur when working with closed equations and not taking into account earlier and later states that the samples will pass.\\

Consider, for example, the case for $L=3$ and the simple dataset consisting of the samples 000, 100, 011, 011, 111.\\

Then we get the following coefficients:\\
$N_{000} = 1/5,\ N_{001} = 0,\ N_{010} = 0,\ N_{011} = 2/5,\ N_{100} = 1/5, N_{101} = 0, N_{110} = 0,\ N_{111} = 1/5$.\\

When we consider only this snapshot for obtaining the proportions $P(i)$ of trajectories in a certain node $i$, we get \\
$P(000) = 1/5, \ P(011) = 2/5,\ P(100) = 1/5,\ P(111) = 1/5 \text{ and } P(001) = P(010) = P(101) = P(110) = 0$.\\

When considering now, for example, the equation for $P(011)$ from \eqref{orig_equations}, then we have
\begin{equation*}
	P(011) = P(010)\cdot a_{010;011} + P(001)\cdot a_{001;011}.
\end{equation*}
\\

Setting in the proportions $P(i)$ from above, we obtain
\begin{equation*}
	\frac{2}{5} = 0\cdot a_{010;011} + 0\cdot a_{001;011} = 0,
\end{equation*}
what is a contradiction.\\

Since sampled data always consists of selective excerpts, such scenarios are entirely possible for many data sets, which renders this simple method unusable. Therefore, it was necessary to include previous and subsequent steps, which resolved this issue.

%
%
\section{Detailed results ovarian cancer data}
\label{appendix_results}

In this appendix, we give the detailed parameter values we obtained via the different methods discussed in Subsection \ref{subsec_ovarian} for the ovarian cancer dataset. Furthermore, also the deviations from zero after evaluating all 28 polynomials at the corresponding parameter values can be found here.
\begin{table}
	\centering
	\caption{Parameter values obtained by the different approaches, based on the ovarian cancer data set \cite{Knutsen_2005} with Feature 1: $8q+$, Feature 2: $3q+$, Feature 3: $5q-$, Feature 4: $4q+$. The $a_{i;j}$ values for HyperLAU, HyperTraPS, and HyperHMM are given by the algorithms directly. The values for the $b_{i;j}$ are obtained substituting these $a_{i;j}$ into the polynomials obtained via HyperALG and minimising the sum of the squares of the remaining polynomials. For HyperALG both the $a_{i;j}$ and the $b_{i;j}$ values are obtained by this minimisation method. All numbers were rounded to four decimal places. \\}
	\label{variable_values}

	\begin{center}
	\begin{tabular}{c|cccccc}
		& HyperLAU & HyperTraPS & HyperHMM& $\text{HyperALG}_1$ & $\text{HyperALG}_2$ & $\text{HyperALG}_3$\\
		\hline
		$a_{0000;1000}$ & 0.5024& 0.4684&0.5013 & 0.5&0.5&0.6652\\
		$a_{0000;0100}$ &0.1398&0.0771& 0.1432 &0.1429&0.1429&0.1466\\
		$a_{0000;0010}$ &0.2157&0.2561& 0.2133 &0.2143&0.2143&0.1175\\
		$a_{0000;0001}$ &0.1421&0.1984&0.1422&0.1428&0.1428&0.0707\\
		$a_{0001;1001}$ &0.4973&0.6125&0.2438&0.3248&0.2323&0\\
		$a_{0001;0101}$ &0.0468&0.081&0.1703&0.1631&0.1904&0.065\\
		$a_{0001;0011}$ &0.4559&0.3065&0.5859&0.5121&0.5773&0.935\\
		$a_{0010;1010}$ &0.21&0.0092&0.0761&0.0928&0.0925&0\\
		$a_{0010;0110}$ &0.052&0.0006&0.2731&0.1689&0.2348&0.1211\\
		$a_{0010;0011}$ &0.738&0.9902&0.6508&0.7383&0.6727&0.8789\\
		$a_{0011;1011}$ &1&0.9538&0.9795&0.8636&0.8622&0\\
		$a_{0011;0111}$ &0&0.0462&0.0205&0.1364&0.1378&1\\
		$a_{0100;1100}$ &0.2425&0.2878&0.7309&0.6182&0.7047&0.4909\\
		$a_{0100;0110}$ &0.5212&0.703&0.1564&0.2602&0.2044&0.3483\\
		$a_{0100;0101}$ &0.2363&0.0092&0.1127&0.1216&0.0909&0.1608\\
		$a_{0101;1101}$ &0.2365&0.309&0.5252&0.5956&0.6762&0.0964\\
		$a_{0101;0111}$ &0.7635&0.691&0.4748&0.4044&0.3238&0.9036\\
		$a_{0110;1110}$ &0.7596&0.1093&0.6199&0.9595&0.9221&0.4012\\
		$a_{0110;0111}$ &0.2404&0.8907&0.3801&0.0405&0.0779&0.5988\\
		$a_{1000;1100}$ &0.8707&0.8249&0.7483&0.7678&0.7398&0.575\\
		$a_{1000;1010}$ &0&0.0733&0.0559&0.0482&0.0468&0.0833\\
		$a_{1000;1001}$ &0.1293&0.1018&0.1958&0.184&0.2134&0.3417\\
		$a_{1001;1101}$ &0&0.908&0.0177&0&0&0.0316\\
		$a_{1001;1011}$ &1&0.092&0.9823&1&1&0.9684\\
		$a_{1010;1110}$ &0.0005&0.0115&0.0249&0.2136&0.3635&0.3799\\
		$a_{1010;1011}$ &0.9995&0.9885&0.9751&0.7864&0.6365&0.6201\\
		$a_{1100;1110}$ &0.558&0.9771&0.5872&0.5363&0.5317&0.5781\\
		$a_{1100;1101}$ &0.442&0.0229&0.4128&0.4637&0.4683&0.4219\\
		$b_{0001;1001}$ &0.5206&0.7082&0.2609&0.3353&0.2372&0\\
		$b_{0001;0101}$ &0.1785&1&0.5996&0.573&0.677&0.1629\\
		$b_{0001;0011}$ &0.2872&0.1874&0.3751&0.3246&0.3639&0.421\\
		$b_{0010;1010}$ &1&0.1257&0.3664&0.4522&0.4586&0\\
		$b_{0010;0110}$ &0.1418&0.0863&0.7221&0.5315&0.6328&0.2178\\
		$b_{0010;0011}$ &0.7128&0.8126&0.6249&0.6754&0.6361&0.579\\
		$b_{0011;1011}$ &0.5543&0.8814&0.5562&0.51&0.5156&0\\
		$b_{0011;0111}$ &0.0054&0.2524&0.0847&0.6099&0.6193&0.7449\\
		$b_{0100;1100}$ &0.0717&0.0527&0.2182&0.187&0.2139&0.1584\\
		$b_{0100;0110}$ &0.8582&0.9137&0.2779&0.4685&0.3672&0.7822\\
		$b_{0100;0101}$ &0.8215&0&0.4004&0.427&0.323&0.8371\\
		$b_{0101;1101}$ &0.0426&0.0192&0.0956&0.1057&0.1174&0.0135\\
		$b_{0101;0111}$ &0.5939&0.1839&0.3521&0.3264&0.258&0.1006\\
		$b_{0110;1110}$ &0.1945&0.0093&0.1502&0.2247&0.2162&0.0846\\
		$b_{0110;0111}$ &0.4005&0.5637&0.5632&0.0637&0.1227&0.1545\\
		$b_{0111;1111}$ &0.0515&0.0878&0.0544&0.0504&0.0504&0.2531\\
		$b_{1000;1100}$ &0.9283&0.9476&0.7818&0.813&0.7861&0.8416\\
		$b_{1000;1010}$ &0&0.8743&0.6336&0.5478&0.5414&1\\
		$b_{1000;1001}$ &0.4794&0.2918&0.7391&0.6647&0.7628&1\\
		$b_{1001;1101}$ &0.0025&0.9123&0.0108&0&0&0.0341\\
		$b_{1001;1011}$ &0.3353&0.0557&0.3335&0.3993&0.4118&0.8941\\
		$b_{1010;1110}$ &0.0014&0&0.0035&0.0277&0.0463&0.0671\\
		$b_{1010;1011}$ &0.1104&0.0629&0.1103&0.0907&0.0726&0.1059\\
		$b_{1011;1111}$ &0.3992&0.3286&0.3914&0.3816&0.3789&0.2358\\
		$b_{1100;1110}$ &0.8041&0.9907&0.8463&0.7476&0.7375&0.8483\\
		$b_{1100;1101}$ &0.9549&0.0685&0.8936&0.8943&0.8826&0.9524\\
		$b_{1101;1111}$ &0.2202&0.1652&0.2216&0.2292&0.2314&0.2014\\
		$b_{1110;1111}$ &0.3291&0.4184&0.3326&0.3388&0.3393&0.3097
	\end{tabular}
	\end{center}
\end{table}

\begin{table}
	\centering
	\caption{Deviations from zero after substituting the values from Table \ref{variable_values} into the 28 polynomials obtained by HyperALG (rounded to four decimal places).\\}
	\label{derivations}

	\begin{center}
	\begin{tabular}{ccccccc}
		HyperLAU & HyperTraPS & HyperHMM &$\text{HyperALG}_1$& $\text{HyperALG}_2$ & $\text{HyperALG}_3$\\
		\hline
		0.0016 &0.0085&0&-0.004&-0.0016&-0.0097\\
		0.0022 &0.0126&0&0&0&-0.0038\\
		0.0005 &0.0089&0&0.0064&0.0009&0.0011\\
		-0.0002 &-0.0097&0&0.0012&0.0015 &0\\
		-0.0009 &-0.0068&0&0.0005&0.0002&0.0006\\
		0.0002 &-0.0037&0&-0.0046&-0.0003&0.0005\\
		-0.0006 &-0.0021&0&0&0&0.0032\\
		0.0016 &0.0031&0&-0.0052&-0.0066&0.0007\\
		0.0004 &0.0049&0&-0.0001&0&0.0049\\
		0.0006 &-0.0007&-0.0001&-0.0006&-0.0008&-0.0011\\
		-0.0002 &0.0002&0&0.0039&0.0045&0\\
		-0.0004 &-0.0021&0&-0.0006&-0.0006&0.0014\\
		0.0005 &0.0008&0&-0.0003&-0.0002&-0.0005\\
		0.0008 &0.0051&0&-0.0026&-0.0002&-0.0004\\
		0.0003 &0.0085&0.0001&0.0001&0&0.002\\
		-0.0006 &0.0002&0&0.0001&0&-0.0008\\
		0.0006 &0.0006&0&-0.0015&-0.0003&0\\
		0 &0.0025&0&0.0001&0&0\\
		0.0002 &0.0086&0&-0.0048&-0.0048&0\\
		0.0001 &0.0031&0&0.0053&0.0054&0.0047\\
		0.0001 &-0.0028&0&0.0002&0.0008&-0.0012\\
		-0.0002 &-0.0039&0&0.0012&0.0015&0\\
		0.0006 &0.0025&0&0&0&-0.0001\\
		0.0004 &-0.0031&0&0.0005&-0.0008&-0.0002\\
		0.0004 &-0.0004&0.0001&-0.0001&-0.0002&0.0019\\
		-0.0002 &-0.0013&0&0.0002&0&0.0111\\
		0.0005 &0.0038&0&0.0059&0.0084&-0.0109\\
		-0.0001 &0.0014&0&-0.0074&-0.0084&-0.0001
	\end{tabular}
	\end{center}
\end{table}


\end{document}